\documentclass[prd,aps,floats,twocolumn,nofootinbib]{revtex4-1}
\usepackage{slashed}
\usepackage{mathtools}
\usepackage{amsfonts}
\usepackage{amssymb}
\usepackage{epsfig}

\usepackage{bm}

\begin{document}

\newcommand{\m}[1]{\mathcal{#1}}
\newcommand{\nn}{\nonumber}
\newcommand{\ph}{\phantom}
\newcommand{\eps}{\epsilon}
\newcommand{\be}{\begin{equation}}
\newcommand{\ee}{\end{equation}}
\newcommand{\bea}{\begin{eqnarray}}
\newcommand{\eea}{\end{eqnarray}}
\newtheorem{conj}{Conjecture}

\newcommand{\plk}{\mathfrak{h}}


\title{
Attractive voids}
\date{}

\author{Raymond Isichei}

\author{Jo\~{a}o Magueijo}
\email{j.magueijo@imperial.ac.uk}
\affiliation{Theoretical Physics Group, The Blackett Laboratory, Imperial College, Prince Consort Rd., London, SW7 2BZ, United Kingdom}

\begin{abstract}
We explore the well know mass deficit/surplus phenomenon in General Relativity to suggest that it could play a part in the dark matter conundrum. Specifically in collapses and condensations of matter associated with negative intrinsic curvature  of the foliation associated with the asymptotic boundary conditions, the external (ADM) mass can vastly exceed the integrated local energy over the internal volume. This can be phrased in terms of a deficit of volume for a given surface area (with respect to zero curvature). We explore the phenomenon in the context of generalizations of the Oppenheimer-Snyder models and other ``cut and paste'' models, the Lemaitre-Bondi-Tolman metric and several others. We produce constructions where the internal object is contracting or expanding, has a life time different from the asymptotic Universe, as well as a volume different than the escavated volume from the Universe. These are purely relativistic constructions and they could play a role in the puzzle of dark matter: attraction without visible or indeed any matter. 


 
\end{abstract}

\maketitle

\section{Introduction}

It has for long been noted that, with suitable definitions and conditions on the space-time, the asymptotic mass of an extended object may differ from the local mass integrated over its volume elements. This ``mass deficit'' pervades classics of literature such as  Zeldovich and Novikov~\cite{Zelbook}, and can be interpreted in a quasi-Newtonian setting as the result of the ``gravitational binding energy'' of the object. It applies to stable astrophysical objects, such as stars, but also to collapsing objects. This can be illustrated by the Oppenheimer-Snyder (OS) toy model of black hole collapse~\cite{OS,MTW}  where an internal piece of a collapsing closed FRW model is collated onto an outer Schwarzchild solution. The matching conditions reveal that the integrated mass in the FRW interior is larger than the Schwarzchild mass parameter (see e.g.~\cite{Misnerstudent,Stephani,Guth}). As this example shows, this effect can transcend quasi-Newtonian physics and be purely relativistic. In the extreme case we could construct an OS model using a near complete FRW 3-sphere for the internal section, generating an infinitesimal outside mass, making the system elusive to detection by the outside world.  This would be a frustrated baby Universe~\cite{BabyU}.


The point of this paper is two-fold. First we aim to generalize the formalism to more complicated settings, specifically making use of the Lemaitre-Tolman-Bondi (LTB) metric~\cite{exactslns,enqvist,Goncalves}. This will allow us to consider a multitude of other scenarios where this phenomenon is present, relevant to gravitational collapse but also to the primordial initial conditions of our Universe. Second, we wish to highlight the fact that in the same way that there can be a mass deficit, there can also be a mass surplus. In extreme situations (converse to the frustrated baby Universe example~\cite{BabyU}) it is possible to generate asymptotic mass even without any local energy density.  It is tempting to relate this feature to the puzzle of dark matter: gravitational pull without visible matter. Perhaps some of the ``dark'' matter is no matter at all. We provide examples of such ``attractive voids'', as in the title of this paper. 

In addition to these two core motivations, we present several examples (and even a general class of solutions) with the property that not only the inferred outside mass, but also its inferred outside volume and age are different from the internal ones. In fact the mass deficit/surplus and the volume mismatches are related. All of these structures show that extrapolation from the outside to the inside may miss the mark. This is a simple point, related to the well know and accepted relativity of time, space and mass, but we believe it has not been properly digested in the context of dark matter studies. 

Could these objects exist? There are several contexts in which these structures could come to exist in the real world. They could have an astrophysical origin in collapses dominated by kinetic energy in a Newtonian setting or, perhaps more importantly, their relativistic equivalent. Such collapses might be better characterized as "implosions'', and complement the usual picture of gravitational collapse. In a different context, these structures could also be part of the primordial Universe, forming counterparts to primordial black holes~\cite{PBH,PBHreview}. They could even be a component of the initial conditions of our Universe.



\section{Mass mismatch in Openheimer-Snyder collapse}\label{SecOS}
The collapsing  phase of a $k=1$ FRW dust model can be used as a toy model for collapse into a black hole. This is the Oppenheimer-Snyder (OS) model~\cite{OS,MTW}, and it can be generalized~\cite{Stephani,Guth} to inside FRW metrics 
with the full set of options $k=0,\pm 1$, as we now review. For $\chi\le \chi_0$ we take:
\begin{equation}\label{FRW}
    ds^2=- d\tau^2+a^2(\tau)(d\chi^2+f^2(\chi) d\Omega^2_2)
\end{equation}
where:
\begin{eqnarray}
    f(\chi)=&\sin\chi&{\rm for}\;k=1\nn\\
    =&\chi&{\rm for}\;k=0\nn\\
    =&\sinh\chi&{\rm for}\;k=-1.\nn
\end{eqnarray}
The solutions to the Friedmann equations for dust are well known. For $k=0,-1$ one takes the branch containing contracting solutions (unlike for $k=1$, these are disconnected from  expanding solutions, a crucial difference).  
The outside metric for all $k$ is  the Schwarzchild metric:
\begin{equation}\label{Schw}
    ds^2=-\left(1-\frac{R_S}{R}\right)dt^2+\frac{dR^2}{1-\frac{R_S}{R}}+R^2d\Omega^2_2
\end{equation}
(where $R_S=2GM$). The matching is performed on a radial geodesic surface, so we consider the radial geodesics of \eqref{Schw}. These satisfy~\cite{MTW,Stephani,Guth}:
\begin{equation}\label{geodSchwarz}
    \left(\frac{dR}{d\tilde \tau}\right)^2+\left(1-\frac{R_S}{R}\right)=E^2,
\end{equation}
with $\tilde\tau$ the proper time and $E$ an integration constant. We must select  the appropriate $E$ to obtain the correct match to the various $k=0,\pm 1$ cases.
The $\chi=\chi_0$ internal geodesic must match a radial outside geodesic for some $R=R(\tilde\tau)$ (with $\tau=\tilde\tau$, since $\tau$ in (\ref{FRW}) is also the proper time of the $\chi=\chi_0$ geodesic). From the form of the solutions for these geodesics ~\cite{MTW,Stephani,Guth} one sees that the correct matching is achieved with:
\begin{eqnarray}
    k=1&\longleftrightarrow&E^2<1\\
     k=-1&\longleftrightarrow&E^2>1\\
      k=0&\longleftrightarrow&E^2=1.
\end{eqnarray}
Hence, the internal closed, open and flat FRW models must match bounded, unbounded and borderline outside geodesics, respectively. More generally we can check all the jumping conditions, reviewed in Appendix ***,  but these do not add any new conditions.

The matching conditions translate into a condition between the Schwarzchild (here also ADM) mass $M$ and $\rho$, $a$ and $\chi_0$~\cite{OS,MTW,Stephani,Guth}:
\begin{eqnarray}
    M=\frac{4\pi}{3}\rho a^3 f^3(\chi_0)=M_\star f^3(\chi_0)
\end{eqnarray}
where we defined
\begin{equation}
    M_\star\equiv \frac{4\pi}{3}\rho a^3
\end{equation}
for later convenience. 
This is to be contrasted with the integrated local mass of the dust inside the FRW model:
\begin{equation}
     M_F=4\pi \rho a^3\int _0^{\chi_0}d\tilde\chi \, f^2(\chi)= 3M_\star g (\chi_0)
\end{equation}
where:
\begin{eqnarray}
    g(\chi)=&\frac{\chi}{2}-\frac{\sin2\chi}{4}&{\rm for}\;k=1\nn\\
    =&\frac{\chi^3}{3} &{\rm for}\;k=0\nn\\
    =&-\frac{\chi}{2}+\frac{\sinh 2\chi}{4}&{\rm for}\;k=-1.\nn
\end{eqnarray}
Except for $k=0$, we have $g\neq f^3/3$, leading to the advertised
mismatch between $M$ and $M_F$.
We see that for $k=1$ ($E^2<1$, i.e. bound external geodesics) we have a mass deficit, $M<M_F$. In a Newtonian picture this is the case where we hide mass in a cloak of binding energy. For $k=-1$ ($E^2>1$, i.e. unbound external geodesics), we have instead a surplus of mass, $M>M_F$. In a Newtonian picture this can be seen as the effect of the kinetic energy present, in addition to the rest mass and gravitational binding energy. But the effect is present even when $\chi_0$ is not small, indeed the larger the $\chi_0$, the bigger the mass mismatch. In extreme cases we therefore have a purely relativistic effect, rendering the Newtonian interpretation irrelevant.  

An extreme case is that of a near-baby Universe~\cite{BabyU}, resulting from $k=1$ and $\chi_0=\pi^-$. Then, $M=0^+$, whilst $M_F=3/2 M_\star$. This is a hidden world, with its matter and volume cloaked from the outside Universe. 
For $k=-1$ and $\chi_0\gg 1$ we have the opposite situation (and there is no limit on $\chi_0$). In the extreme case, the internal world is empty: a Milne Universe; yet the outside world sees matter. 

\subsection{Quantifying the effect}

To quantify this effect in terms of more tangible quantities, we must work out how much of the FRW we must cover in a given OS model, that is the value of $\chi_0$ needed for a given set of parameters defining the object. For the $k=1$ model this is usually done by examining the relation between mass and density at the standard initial condition~\cite{MTW} of turnaround $\dot a=0$. Such a point does not exist for $k=-1$. As a generalization we can extract $\chi_0$ from: 
\begin{equation}
    f^3(\chi_0)=\frac{M}{M_\star}.
\end{equation}
Note that both $M$ and $M_\star$ are constants. Up to a numerical factor, $M_\star$ is the first integral of the conservation equation $m=\rho a^3$. 
We can work it out from $\rho$ and $\Omega$, where as usual $\Omega=\rho/\rho_c$, and $\rho_c$ is the value of $\rho$ obtained by setting $k=0$ in the Friedman equation, keeping the other variables ($a$ and $\dot a$) fixed.
If we set $k=\pm 1$, then $a$ has units of length and satisfies:
\begin{equation}
    a^2=\frac{\alpha}{\rho}\frac{\Omega}{|\Omega-1|},
\end{equation}
where $\alpha=3/(8\pi G)$ is proportional to the reduced Planck mass squared. 
After some manipulations we arrive at:
\begin{equation}
    M_\star(\rho,\Omega)=\frac{4\pi}{3}\frac{\alpha^{3/2}}{\sqrt{\rho}}\left(\frac{\Omega}{|\Omega -1|}\right)^{3/2}
\end{equation}
With this formula we can evaluate $\chi_0$ from the mass, density and the value of $\Omega$ at any point, that is:
\begin{equation}
    \chi_0=\chi_0(M,\rho,\Omega).
\end{equation}
For the standard OS ($k=1$) at the turnaround point we have $\Omega=\infty$ and:
\begin{equation}
    M_\star(\rho,\infty)=\frac{4\pi}{3}\frac{\alpha^{3/2}}{\sqrt{\rho}}. 
\end{equation}

\subsection{Geometrical interpretation}
\label{surpdef}
The phenomenon we illustrated within OS follows the pattern
    {\it The outside gravitational mass of a homogeneous spherical dust region with a given surface area equals the integrated stress-energy mass over a volume with the same surface area but flat. } 
The mass deficit and excess can then be understood geometrically as the result of this volume being larger or smaller depending on whether we have positive or negative intrinsic curvature. 

\begin{figure}[ht]
\begin{tabular}{c}
\includegraphics[trim=0cm 3cm 4cm 3cm, 
scale=0.6]{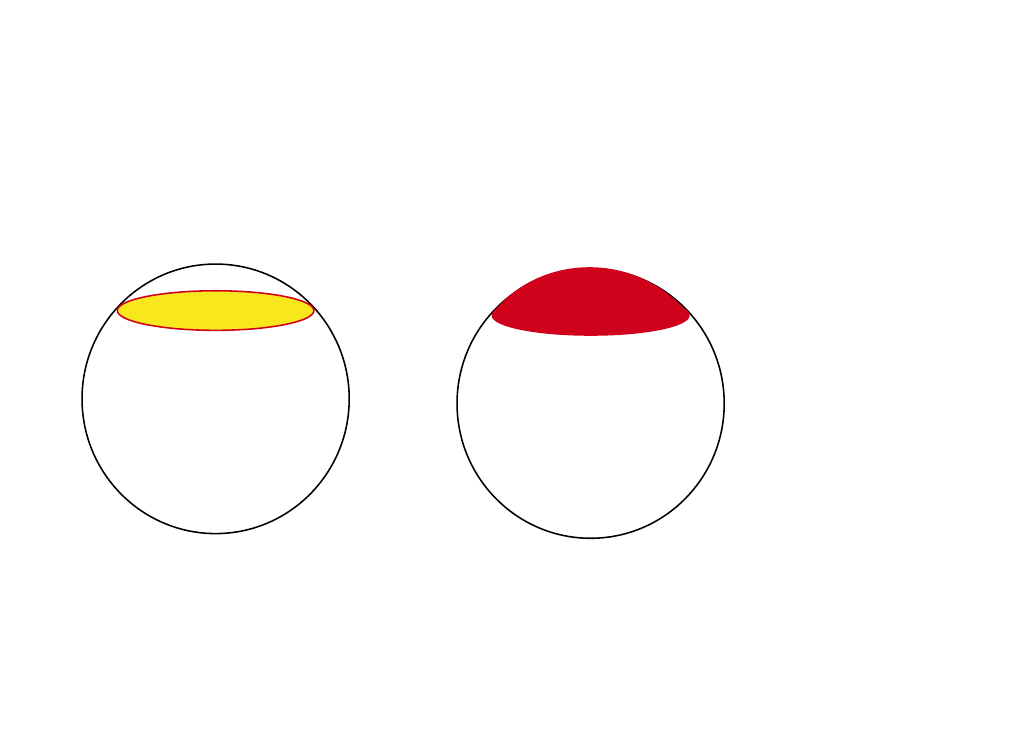}
\end{tabular}
\caption{The mass deficit for a $k=1$ collapse can be understood as a volume excess (red region) for the same surface area (red line) with respect to euclidean space (yellow region). } \label{Volexcess}
\end{figure}

\begin{figure}[ht]
\begin{tabular}{c}
\includegraphics[trim=2cm 3cm 4cm 3cm, 
scale=0.5]{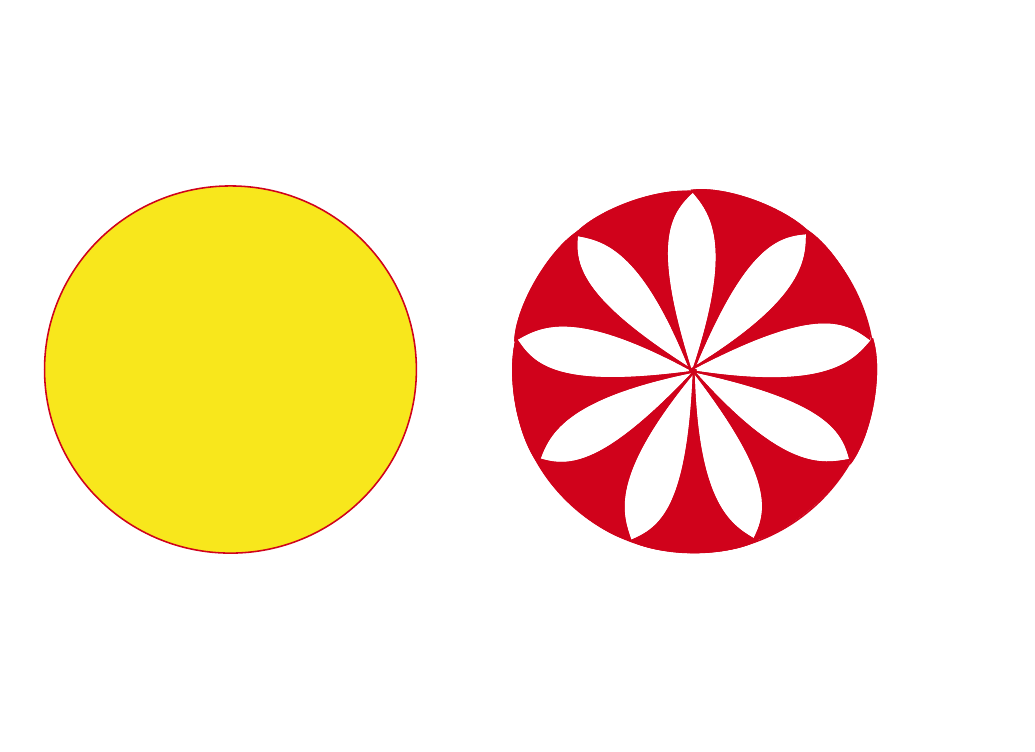}
\end{tabular}
\caption{The equivalent situation for $k=-1$ can be illustrated using the fact that geodesics shooting off from the pole diverge, instead of converging as in the $k=1$ case. Filling in the space between them in red (right picture) leaves a volume deficit  (the regions in white) with  respect to an euclidean space with the same boundary area (yellow region). This translates into a mass surplus. }
\label{Voldeficit}
\end{figure}

This is illustrated in Figs.~\ref{Volexcess}/\ref{Voldeficit} for positive/negative curvature. In the second case we use the fact that geodesics shooting away from the ``pole'' converge for positive curvature, but diverge for negative curvature, in the sense that the angle between them increases. This allows us to represent the volume deficit (translating into a mass surplus) for the $k=-1$ case. These figures are the converse of those in~\cite{carroll}.

We could quantify these interpretation in terms of the dimensionless ratio between volume and area:
\begin{align}
     \frac{V}{A^{3/2}}=\frac{1}{2\sqrt{\pi}r^3}\int^r_0\frac{d\tilde r \tilde r^2}{1-k\tilde r^2}  
\end{align}
with a reference flat value of $\frac{1}{6\sqrt{\pi}}$. 
The Milne limit is interesting in that in areal coordinates $V\sim 4\pi r$, so this ratio goes like $1/r$. Asymptotically it can be thought of as a finite area enclosing a zero volume. 


This has an observational meaning. Suppose that we take a near baby Universe~\cite{BabyU} full of stars. The brightness for the surface area will be very large. The Keplerian mass will be negligible. Same in reverse with a $k=-1$ void. Perfect darkness (zero brightness) but large Keplerian mass. Brightness is flux, so energy per unit area. The energy comes form the volume but passes through the surface area. So it is sensitive to  this area/volume deficit/surplus.

\section{Critical review of  the foundations of LTB}\label{LTBreview}
The Lemaitre-Tolman-Bondi (LTB) formalism~\cite{exactslns,enqvist,Goncalves} starts from the metric\footnote{Of the large number of notations used in the literature we elected~\cite{enqvist}.}:
\begin{equation}\label{pre-LTB}
    ds^2=-dt^2+X^2(r,t) dr^2+A^2(r,t)d\Omega^2,
\end{equation}
before it is brought to the LTB form. At this stage only spherical symmetry has been assumed. 

If $A'=0$ at isolated points\footnote{We do not consider the case where $A'=0$ in an open set; see Theorem 15.7 of~\cite{exactslns}. None of what follows applies to that case.} it may happen that several $r$ have the same area $A$. In the literature this is usually associated with physical ``shell crossing''~\cite{exactslns,Goncalves}, implying that points with the same $A$ are at the same location. We contend that this is the wrong interpretation within LTB (it leads to a different formalism). 
Metric \eqref{pre-LTB} implies a foliation with a normal $n_\mu=(1,\mathbf{0})$ which is a 
geodesic 4-vector\footnote{This can be proved for any metric where $N=N(t)$ (the fact that $N^i=0$, valid here, is not needed). If $n_\mu=-N dt$, then $n_\mu=\partial_\mu \tilde t$ is locally equivalent to $\partial_i N=0$ (so that $dn=0$). But then (since partials and nablas are equivalent in the exterior derivative) $\nabla_\mu n_\nu=\nabla_\nu n_\mu$, so combined with $n_\mu n^\mu=-1$ we have $n^\mu\nabla_\mu n^\alpha=0$.}. It is then {\it assumed} that $n^\mu$ is also the (geodesic) velocity 4-vector of the total dust matter at each point: this is 
why these coordinates are dubbed ``comoving'' coordinates. Then the stress energy tensor in these coordinates only has a $00$ component, since:
\begin{align}
    T_{\mu \nu}&=\rho \, n_\mu n_\nu .
\end{align}
If shells were to cross at the same physical point, then the stress energy tensor would instead have the form:
\begin{equation}
    T^\mu_{\;\nu}=\rho_1 \, n_\mu n_\nu +\rho_2 u_\mu u_\nu 
\end{equation}
with $u^\mu\neq n^\mu$ in general, so that $T_{\mu\nu}$  would have components other than 00. In such a situation, 
the Einstein equations would not be the ones found in the LTB formalism before the metric is brought to its final form, viz~\cite{enqvist}:
\begin{widetext}
\begin{align}
-2\frac{A''}{AX^2}+2\frac{A'X'}{AX^3}+2\frac{\dot{X}\dot{A}}{AX}+\frac{1}{A^2}+\left(\frac{\dot{A}}{A}\right)^2
-\left(\frac{A'}{AX}\right)^2 &= 8\pi G\rho\label{ham}\\
\dot{A}'-A'\frac{\dot{X}}{X}&=0\label{mom}\\
2\frac{\ddot{A}}{A}+\frac{1}{A^2}+\left(\frac{\dot{A}}{A}\right)^2-\left(\frac{A'}{AX}\right)^2 &=0 \label{dyna1}\\
-\frac{A''}{AX^2}+\frac{\ddot{A}}{A}+\frac{\dot{A}}{A}\frac{\dot{X}}{X}+\frac{A'X'}{AX^3}+\frac{\ddot{X}}{X}
&= 0.\label{dyna2}
\end{align}
\end{widetext}
The Hamiltonian content of \eqref{ham}-\eqref{dyna2} can be read off easily. Eq.\eqref{ham} is the $00$ component of Einstein's equations, and so represents the Hamiltonian constraint. Eq.\eqref{mom} is their $0i$ component, and so represents the diffeomorphism or  momentum constraint. The last two equations are the $rr$ and $\theta\theta$ (or $\phi\phi$) components, making up the dynamical or Hamilton's equations.

If there were physical shell crossing, these equations would contain an extra source with non-$00$ components. Specifically, they would have a source term in the momentum constraint, essential for bringing the metric to the LTB form, as we will presently show. For LTB to follow, we must assume no matter currents contribute to the momentum constraint.
Thus, the LTB formalism requires more than spherical symmetry. It requires a specific type of dust source which precludes shell crossing. It requires that  $r$ be a good coordinate, so that points with different $r$ and the same $A$ are different points in space and no shell-crossing occurs. This is not that abstruse: in fact the FRW $k=1$ solution expressed in LTB coordinates offers the simplest example (as we will see in Section~\ref{SecFRW}): as the radial coordinate increases the area first increases, then past the equator it decreases.


\section{LTB metric, equations and solutions}

With this matter clarified the usual LTB derivations do follow through.  
In points where $A'\neq 0$, the source-free momentum constraint, Eq.\eqref{mom}, implies  $X=C(r)A'$. Writing $C(r)=1/\sqrt{1-K(r)}$ with $K(r)<1$, reduces the metric to the LTB form:
\begin{equation}\label{LTB}
    ds^2=-dt^2+\frac{A'^2}{1-K(r)}dr^2+A^2d\Omega^2.
\end{equation}
The remaining Einstein equations can then be brought to:
\begin{eqnarray}
    \frac{\dot A^2}{A^2}&=&\frac{F(r)}{A^3}-\frac{K}{A^2}\label{LTBF1}\\
    \frac{2}{3}\frac{\ddot A}{A}+\frac{1}{3}\frac{\ddot A'}{A'}&=&-\frac{4\pi G}{3}\rho,\label{LTBF2}
\end{eqnarray}
where 
\begin{equation}\label{Fprime}
    \frac{F'}{A'A^2}=8\pi G\rho.
\end{equation}
These can be condensed further into: 
\begin{eqnarray}
  \frac{\dot A^2}{A^2}&=&\frac{F(r)}{A^3}-\frac{K}{A^2}\\
  \dot F&=&0\label{LTBC1}\\
  \dot K&=&0\label{LTBC2}
\end{eqnarray}
with constraint:
\begin{equation}\label{Fprime}
   F'=8\pi G\rho A'A^2.
\end{equation}
Hence the time independent functions $F(r)$ and $K(r)$ can be seen as the inputs. 
If we want to read off the density from the constant function $F(r)$ (which parameterizes the solutions) we can reqrite \eqref{Fprime} as:
\begin{equation}\label{density}
    \rho=\frac{F'}{8\pi G A'A^2}.
\end{equation}
It implies the conservation law:
\begin{equation}\label{cons}
    \dot \rho + \left(2\frac{\dot A}{A}+\frac{\dot A'}{A'}\right)\rho=0.
\end{equation}
Obviously if $\rho\ge 0$ and finite, we must have that $F'/A'>0$, that is $F'$ and $A'$ must have the same sign. If $F'=0$ then we are in vacuum and $A'$ can have any sign. In points where $A'=0$ we must also have $F'=0$.

The solutions to this system in regions where $A'\neq 0$ are well known, and separate for each $r$. For $K(r)>0$:
\begin{align}
    A&=\frac{F}{2K}(1-\cos\eta)\label{sol1}\\
    t-t_0&=\pm \frac{F}{2K^{3/2}}(\eta-\sin\eta);
\end{align}
for $K(r)<0$:
\begin{align}
    A&=\frac{F}{2|K|}(\cosh\eta-1)\\
    t-t_0&=\pm \frac{F}{2|K|^{3/2}}(\sinh\eta-\eta);
\end{align}
and for $K(r)=0$:
\begin{equation}
t-t_0=\pm \frac{2}{3}\frac{A^{3/2}}{F^{1/2}}
\end{equation}
or
\begin{equation}
    A=\left(\frac{9F}{4}\right)^{1/3}|t-t_0|^{2/3}\label{sollast}
\end{equation}
where $t_0=t_0(r)$ is a free function. The $K\neq 0$ solutions reduce to the $K=0$ ones when $\eta\ll 1$. The limit $\eta\rightarrow\infty$ of $K=-1$ is:
\begin{equation}\label{milne}
    A=\sqrt{|K|}(t-t_0(r))
\end{equation}
for any $F(r)$. This is  equivalent to setting to zero the mass term in \eqref{LTBF1} (or to assuming $F\ll |K|A$), so this is a good model for void. If $t_0(r)$ is constant this limit reduces to the Milne Universe.

Some comments on these solutions:

\begin{itemize}
    \item It may seem that  there are 3 independent free functions ($F$, $K$ and $t_0$), but $r$ can be redefined by an arbitrary monotonic function, 
    so in fact there are only 2 independent ones. 
    \item   
    Beware if initial conditions are given in terms of the density, since $A$ appears in its expression (cf. \eqref{density}) which in turn contains $F$ within these solutions. Reading off the appropriate solution may then be non-trivial. 
\end{itemize}

\section{Generating solutions by cutting an pasting}\label{Sec.jump}
The fact that these solutions seemingly decouple in $r$ does not mean that the different shells are entirely decoupled, since the functions $F(r)$ and $K(r)$ cannot be completely arbitrary given the continuity of space. In turn, this affects the process of cutting and pasting  solutions, and can be phrased as jumping conditions. These can be derived directly from the Einstein equations, or appealing to well-known formalisms for gluing solutions across hypersurfaces. 

We backtrack to \eqref{ham}-\eqref{dyna2} and examine the impact of jumping conditions on their left hand side, seeking the implications of avoiding a delta function at a given gluing $r$ (and so the need to introduce a shell of matter in their right hand side). Assuming first $A'\neq 0$ (and also obviously $A\neq 0$), 
we see that absence of delta functions in Eq.\eqref{dyna1} (or \eqref{mom}) implies that around any point $r$:
\begin{align}\label{nojumpA}
    A(r_-)=A(r_+). 
\end{align}
Eq.\eqref{ham} (or \eqref{dyna2}) then shows that $A'$ and $X$ can jump, but only keeping their time independent proportionality factor $C(r)$ fixed. This implies that
\begin{align}\label{nojumpK}
    K(r_-)=K(r_+). 
\end{align}
Other than this, $K$ can change sign or approach 1, its upper bound, at a gluing point. These are the two conditions for gluing at points where $A'\neq 0$.  

In contrast, at isolated points $r=r_\star$ where $A'=0$, Eq.~\eqref{mom} only implies $\dot A'=0$, so such points remain extremes of $A(r)$ at all times. Of the other jumping conditions:
\begin{equation}\label{Astartjump}
    A(r_{\star +})=A(r_{\star -})
\end{equation}
remains the same, but \eqref{ham} (or \eqref{dyna2}) now only implies that $A'$ cannot jump
\begin{equation}\label{Astartjump}
    A'(r_{\star +})=A'(r_{\star -})=0
\end{equation}
whilst $X$ can, so $K$ could jump at these points. This is the opposite of what happens for points with $A'\neq 0$. 

We add two important comments to this conclusion. First, we note that, as can be seen in Appendix~\ref{israel}, these conditions can be replicated appealing to the Israel formalism for cutting and pasting. Second, we stress that there are different variations for these jumping conditions, as critically reviewed in~\cite{bonner}. The stronger versions are deemed to be physically surplus to requirement, and will not be used in this paper.   

(Parenthetically, we note that our cutting and pasting approach is almost orthogonal to that seeking a smooth transition between astrophysical and cosmological structures; eg~\cite{baker,Bolejko,rivers}.)

\section{Critical review of basic solutions}
We first illustrate some of the points made with well known solutions. 
\subsection{FRW within LTB}\label{SecFRW}
The FRW solutions can be obtained by choosing:
\begin{align} 
    F&=\frac{8\pi G}{3}m  r^3\label{idF}\\
    K&=kr^2\label{idK}\\
    t_0&=0\label{t0FRW}
\end{align}
where $k=0,\pm 1$. Then we can recognize the usual FRW metric (and its solutions in \eqref{sol1}-\eqref{sollast}) written in terms of a comoving areal coordinate, with identifications:
\begin{align}
    A&=a(t) r\label{idA}\\
    \rho(t) &= \frac{m}{a^3}.
\end{align}
In terms of these variables, \eqref{LTBF1} and \eqref{LTBF2} reduce to the Friedman and Raychaudhuri equation, and \eqref{cons} to the FRW conservation equation for dust. For latter convenience we can define the ``Schwarzchild radius'' of the FRW Universe as $r_S=2Gm$ (in analogy with $R_S=2GM$). 
Then 
\begin{equation}
    F=r_SV_0(r)
\end{equation}
with $V_0$ the comoving Euclidean ($k=0$) volume for a given area indexed by $r$.  

\begin{figure}[ht]
\begin{tabular}{c}
\includegraphics[trim=0cm 1.5cm 8cm 0cm, 
scale=0.8]{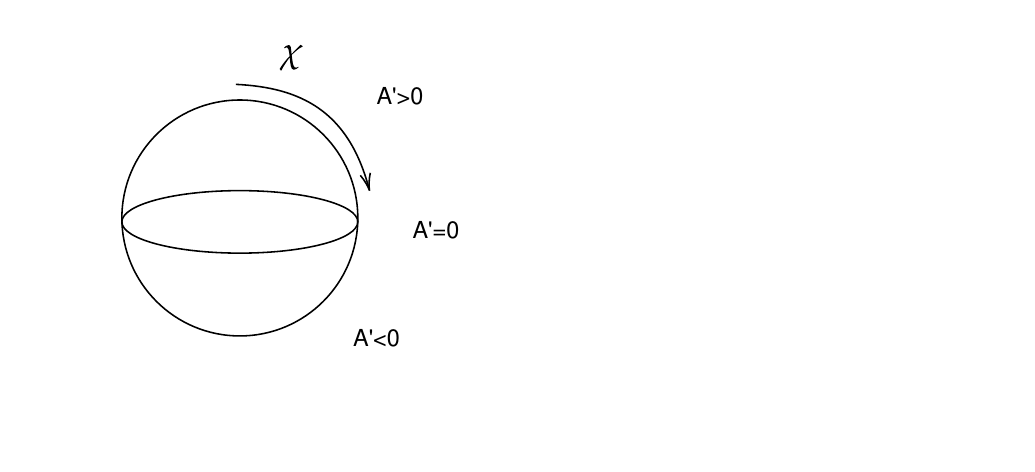}
\end{tabular}
\caption{Simplest example of an LTB solution where $A'=0$ at an isolated point, with $A'$ changing sign, without shell-crossing. } \label{Figsphere}
\end{figure}

We can locally redefine $r$ at will, and a possibility is:
\begin{equation}
    r=\sin\chi
\end{equation}
where $\chi$ is the proper distance to the center/pole and also a latitude angle in the case $k=1$. Indeed this coordinate is needed to cover the whole sphere in the $k=1$ case.  We have $K=r^2$ so $K$ touches 1. To cover the whole sphere we need to 
use $\chi$. Then $A'=0$ at $\chi=\pi/2$, and there are points with different $\chi$ and the same $A$. This clearly does not signal shell crossing, but simply a manifold where the area does not increase monotonically with the radius, as obviously is the case of the 3-sphere.  This is an obvious point, illustrated in Fig.~\ref{Figsphere}, but it helps to understand more complicated solutions later. 

\subsection{Schwarzchild space time variations}\label{SecSchwarz}

Lines of constant LTB spatial coordinates are geodesics, so setting $F$ to constant
\begin{equation}
    F=2GM=R_S
\end{equation}
in \eqref{sol1}-\eqref{sollast} must produce portions of Schwarzchild space-time in timelike geodesic coordinates, each radial geodesic indexed by constant (comoving) $r,\theta,\phi$. These can be incoming or outgoing, depending on the sign choice in \eqref{sol1}-\eqref{sollast}. They can be bounded, unbounded or borderline, corresponding to the various $K(r)$. $t_0(r)$ selects the initial condition for each geodesic, and can be $r$ dependent. Different choices cover different partial patches of the Kruskal manifold and these coordinates can be seen as the time-like equivalent of the Eddington-Finkelstein null coordinates. 

For $K=0$, $t_0$ must depend on $r$, or else $A'=0$ everywhere. Once this is satisfied any monotonic $t_0(r)$ will do. Setting $t_0=r$ corresponds to Lemaitre coordinates~\cite{exactslns}. The radial Schwarzchild coordinate, $R$, being an areal coordinate, relates to the LTB $r$ and $t$ via: 
\begin{equation}\label{rho-A}
    R=A(r,t)
\end{equation}
(since $A^2$ is the areal factor of \eqref{LTB}).
Hence for $K=0$ the horizons are at:
\begin{equation}
    \pm(t-t_0(r))=\frac{2}{3}R_S
\end{equation}
and the singularity is at $t=t_0(r)$. 

For $K\neq 0$ we have significantly more choice
(this is sometime called Novikov coordinates). Comparison with the solutions to \eqref{geodSchwarz}, bearing in mind \eqref{rho-A}, implies:
\begin{align}
    t-t_0(r)&=\tau\\
    K(r)&=1-E^2
\end{align}
Hence $t_0(r)$ is the conventional zero of the proper time of these geodesics, and $K(r)$ the choice for their ``total energy''. Considerable ambiguity is now present.

The choice $K(r)=k/r$ does not lead to the Hubble radial gauge, which in general does not exist (cf.~\cite{Kopinski}). One exception is $k=-1$ and $\eta\gg 1$ which naturally touches Milne (both being empty spaces), so that $K=kr^2$ with $t_0=0$ still is a Hubble gauge. Better find the equivalent of the Lemaitre gauge for when $\eta\ll 1$ and the solutions reduce to $K=0$. This is achieved with:
\begin{align}
    K(r)&=K_0\\
    t_0&=r.
\end{align}

We can also change the sign of $K$ as long as the geodesics do not cross: for outgoing charts this means we can go from $K>0$ to $K<0$, but not the other way, and the opposite for incoming charts.

\subsection{OS and its variations, and vacuoles}
The OS solution and its variations, as well as the similar vacuole solutions (see, e.g.~\cite{Stephani}) can be found within \eqref{sol1}-\eqref{sollast} subject to constraints \eqref{nojumpA} and \eqref{nojumpK}, and exceptions at $A'=0$ points at the end of Section~\ref{LTBreview}. 

For OS-like solutions, at $r<r_0$ we pick an FRW solution, for example with choices \eqref{idF}-\eqref{t0FRW}. Traditionally one takes contracting solutions (with any $k$) but we can also consider expanding ones (modelling explosions rather than collapses). For $r>r_0$ we need a Schwarzchild representation (among those in Section~\ref{SecSchwarz}) satisfying:
\begin{align}
 F(r) =F(r_0)&=\frac{8\pi G}{3}m  r_0^3\label{matchFs}\\
    K(r_0) &=kr_0^2\\
    t(r_0) & =0
 \end{align}
and with the same of sign $t$ at $r_0$ (i.e. the same expanding/outgoing or contracting/incoming nature). With these conditions, $A_+= A_-$ and $K_+=K_-$ at all times.

This still leaves plenty of choice (which can be exploited to link to outer solutions). A possibility is:
\begin{align}
    K&=k r_0^2\\
    t_0&=r-r_0
\end{align}
which reduces to the Lemaitre coordinates when $\eta\ll 1$, i.e. curvature can be ignored. Other possibilities will be explored later. 

The same applies in reverse (Schwarzchild for $r<r_0$, FRW for $r>r_0$), as is the case of a vacuole. 

Thus the matching conditions found in Sec.~\ref{SecOS} (as well as those for the similar vacuole solution; see~\cite{Stephani}) appear naturally, together with the concept of mass deficit/surplus: \eqref{matchFs} implies 
\begin{equation}
     M=\frac{4\pi}{3}\rho a^3 r_0^3 =\rho a^3 V_0(r_0)
\end{equation}
with $V_0$ the comoving Euclidean ($k=0$) volume for the area indexed by areal coordinate $r_0$. The same issues happen in vacuoles where the mass must be the escavated FRW mass corresponding to a $k=0$ FRW Universe, regardless of the actual value of $k$.

\section{Attractive dark voids}

We are now in position to present examples of a large class of solutions in which the mass surplus effect is in evidence. 

\subsection{ OS-FRW combinations}
A vacuole within a FRW Universe containing an OS-like structure in the middle can be obtained combining the matching conditions of the previous subsections. It then all boils down to finding a Schwarzchild chart that links the inner and the outer solutions. We assume there are no mass shells, so $F$ cannot jump, implying:
\begin{align}
    r<r_1 \implies &F=\frac{4\pi}{3}r^3 r_{S1}\nn\\
    r_1<r<r_2 \implies &F= r_{S2} \nn\\
     r>r_2 \implies &F=\frac{4\pi}{3}r^3 r_{S3}\label{Fvacuole}
\end{align}
with:
\begin{equation}
    r_{S2}=\frac{4\pi}{3}r_1^3 r_{S1}=\frac{4\pi}{3}r_2^3 r_{S3}
\end{equation}
which requires:
\begin{equation}
    \frac{r_{S1}}{r_{S3}}=\frac{m_{in}}{m_{out}}=\left(\frac{r_2}{r_1}\right)^3.\label{ratioms}
\end{equation}
In addition, $K(r)$ cannot jump, but we can choose any function interpolating between $K(r_1)=k_1 r_1^2$ and $K(r_2)=k_2 r_2^2$ in the region $r_1<r<r_2$, where $k_1$ and $k_2$ are the normalized curvatures in the inner and outer FRW models (and they all lead to equivalent models). If the inner FRW is contracting and the outer one is expanding, we may need a designer $K(r)$ and $t_0(r)$ to allow the geodesics in $r_1<r<r_2$ to change from incoming to outgoing. 

A large class of solutions therefore follows, with different values of $k_1$ and $k_2$, and with the inner FRW contracting or expanding (we assume the outer FRW is expanding). In addition, we can assume the same bang time $t_0$ for both FRW solutions (so that they have the same age), or not. 

\subsection{A simple example}
A simple example results from the choice:
\begin{align}
    r<r_1 \implies &K=k r^2\\
    r_1<r<r_2 \implies &K=k r_1^2\frac{r_2-r}{r_2-r_1} \\
     r>r_2 \implies &K=0
\end{align}
picking expanding solutions for both FRWs, with the same $t_0$. We focus on the $k=-1$ because it is the relevant one for a mass surplus. The two FRWs are connected by a Schwarzchild buffer zone that can be read off from solutions \eqref{sol1}-\eqref{sollast} choosing the + branch, and setting $t_0(r)=0$ throughout:
\begin{align}
    A&=\frac{r_{S2}(r_2-r_1)}{2r_1^2(r_2-r)}(\cosh\eta-1)\\
    t&=\frac{r_{S2}(r_2-r_1)}{2r_1^2(r_2-r)}   
    (\sinh\eta-\eta);
\end{align}
(note that at $r\approx r_{2-}$ we can take the $\eta\rightarrow 0$ limit). With this choice we satisfy the matching conditions (no jumps in $A$ and $K$) at both gluing points.

The inside/outside solutions are just FRW spaces which touch as we approach the Big Bang\footnote{Proof: find the $\eta\rightarrow 0$ limit of $K\neq 0$ and use \eqref{ratioms}.}. There is a mass surplus in the Schwarzchild region, with respect to the integrated mass in the inside expanding region. As we have seen this results from from a volume deficit inside the sphere bounded by $r=r_1$. We now find a complementary effect to this on the outside. The area at $r=r_2$ is bigger than the area $r=r_1$, with $r_2>r_2$. In the extreme case, these large volumes could be completely hidden from the outside world except for their mass.

\begin{figure}[ht]
\begin{tabular}{c}
\includegraphics[trim=0cm 0cm 0cm 0cm, 
scale=0.6]{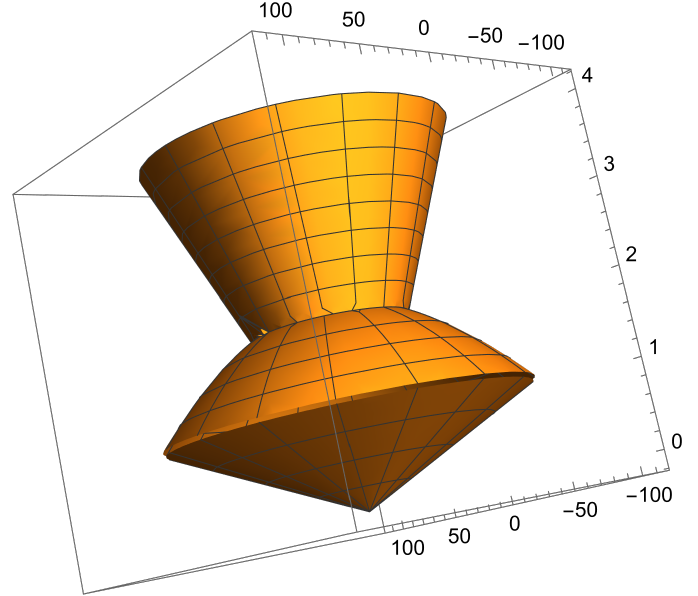}
\end{tabular}
\caption{An expanding inner $k=-1$ Universe which has outgrown the cavity size in the outer $k=0$ Universe. A large internal volume could therefore be hidden in side this structure. }
\label{Fig:hidden}
\end{figure}

Note that this strange property follows from that $A'$ can switch sign, just as it does for a pure $k=1$ FRW model. Here $A'>0$ for the inner and outer expandin FRW, but $A'<0$ in the Schwarzchild buffer. This is indeed what happens here, as depicted by the embedding in Fig.~\ref{Fig:hidden}. The fact that a jump in $A'$ is possible goes back to the issues discussed in~\cite{bonner}, and in Section~\ref{Sec.jump}.

Could such objects be part of the initial conditions of our Universe, and be about and be part of dark matter?

\subsection{Asynchronous Big Bang non-homogeneous generalizations}\label{async}

Finally there could also be a mismatch between the lifetime of the inner and the outer structures, as measured in the synchronous gauge (i.e. by the proper time of the geodesic observers attached to the LTB coordinate frame). 

Note that in the solutions above we could make the Big Bang asynchronous in either inner and outer solution, replacing \eqref{t0FRW} by $t_0=t_0(r)$ and keeping the other 2 choices. Asynchronous Big Bangs have been discussed ($t_0$ is usually called the ``Bang time''). We could also keep $t_0$ constant in the inner and outer FRW but they could take different values.

\subsubsection{Age mismatch effect: Asynchronous Big Bang with an empty buffer}

In the simplest case we can have two homogeneous solutions with different Bang times linked by vacuum. A large class of solutions can be generated this way; here we present a simple example. The point is that generally the inner solution may be larger than the excavated hole of the outer region, leading to situations similar to that depicted in Fig.~\ref{Fig:hidden}.

Consider, for example, the solution obtained from  \eqref{sol1}-\eqref{sollast}  by setting $K=0$ everywhere, choosing $F$ as in \eqref{Fvacuole}, and setting:
\begin{align}
      r<r_1 \implies &t_0^{in}=0\\
    r_1<r<r_2 \implies &  t_0(r)=t_0^{out}\frac{r-r_1}{r_2-r_1}   \\
     r>r_2 \implies &t_0^{out}\neq 0.
\end{align}
This solution satisfies the jumping conditions described in Section~\ref{Sec.jump}. Admittedly $A'$ is discontinuous but that is OK, unless we take seriously the physically unmotivated stronger condition of the 3 classes described in~\cite{bonner}.

If $t_0^{out}=0$ we obviously would have $A_-=A_+$ and $A'=0$ in the open set $r_1<r<r_2$, disqualified by our assumptions. This is just a standard FRW with a fictitious and unnecessary buffer region. If $t_0^{out}>0$ then the inner region is older and larger than the the volume escavated form the outer FRW Universe, and the Schwarzchild like buffer region will be of the same shape as Fig.~\ref{Fig:hidden}. A large volume and past life may then be hidden inside a small hole in a much younger Universe. (The case $t_0^{out}<0$  leads to a more standard vacuole.)

\subsubsection{More general solutions}
More generally $t_0$ might vary with $r$ even within the inner and outer non-vacuum solution, in which case they are no longer pieces of FRW models. Indeed replacing \eqref{t0FRW} by $t_0=t_0(r)$ and keeping \eqref{idF}
and \eqref{idK} implies that $A$ is no longer proportional to $r$ 
and this implies inhomogeneous solutions via~\eqref{density}. Homogeneity implies $F\propto A^3$ at constant time, which implies $t_0'=0$ (by inspection of solutions). For example, for $K=0$, introducing a $t_0'\neq 0$, we have:
\begin{equation}
    \rho=\frac{1}{6\pi G(t-t_0(r))^2\left(1-\frac{2rt_0'(r)}{3(t-t0(r))}\right)}
\end{equation}

We could also keep \eqref{idF} (instead of introducing a vacuum buffer, as in the previous subsection), 
make $t(r)$ constant except for a region linking different Big Bang times $t_0$, thereby we linking two homogeneous FRW models with different uniform densities with a buffer region which is not uniform. 

In general, the singularity time ($\rho\rightarrow \infty$) is the Bang time $t_0$ even if $t_0'(r)\neq 0$.  However, near the singularity, $\rho<0$ if $t_0'>0$, that is, with positive energy we should require the inner shells to have their BB retarded rather than advanced.

We can find Milne-like equivalents from \eqref{milne}. With a choice for $r$ leading to\eqref{idK} we have
\begin{equation}
     A=\pm r(t-t_0(r))
\end{equation}
so that \eqref{idF} and \eqref{density} imply the profile for the density (assumed negligible with respect to the curvature):
\begin{equation}
    \rho=\frac{m}{|t-t_0|^3\left(1-\frac{rt_0'}{|t-t_0|}\right)}
\end{equation}
This solution could be used in variations of all the  mass surplus situations we have examined 


\section{Generalization of the argument for mass surplus and deficit}
With suitable adaptations, we can generalize the argument in Section~\ref{surpdef} to non-homogeneous situations within the LTB framework. The mass felt outside a shell with label $r$ (and not just the asymptotic mass, assuming asymptotic flatness) is given by:
\begin{equation}
    M(r)=\frac{F}{2G}
    =4\pi\int dr\, A'A^2\rho=\int d^3 x\,  \sqrt{h_0}\rho,
\end{equation}
where $h_0$ is the determinant of the spatial metric assuming $K=0$, and 
where we used $A'>0$ in the last step. If $A'$ changes sign and $r(A)$ is multivalued, then we have to take this into account computing the integral: 
\begin{equation}
    M=4\pi\sum _i \int dA A^2\rho_i(A)=\sum_i {\rm sgn}(A')\int d^3 x\,  \sqrt{h_0}\rho,\nn
\end{equation}
(with $i$ indexing all the branches). With these modifications, the mass felt outside the body is still that obtained by filling the boundary with area $A$ with a surface with $K=0$ (taking the possibility that different branches may connect $A=0$ and the given $A$. 

In contrast the integrated local mass is:
\begin{equation}
    M_{L}=\int d^3 x\,  \sqrt{h}\rho
    =4\pi\int dr\, \frac{|A'|A^2}{\sqrt{1-K}}\rho.
\end{equation}
which generalizes $M_F$. The mass mismatch formula is thus generalized. 

If the space is asymptotically flat, then $M$ is the ADM mass.

\section{Conclusions}

In this paper we constructed a number of solutions which generalize the well know mass deficit/surplus effect in various ways. In particular, we found that not only the inferred outside mass may be different to the integrated inside mass, but the volume and lifetime may be different. We can thus produce objects converse to quasi-baby Universes~\cite{BabyU} (which have no asymptotic mass): attractive voids, which have asymptotic mass with minimal internal integrated mass. Such objects, have a deficit of internal volume for a given surface area. They can also have very different ages and that of the asymptotic Universe. 

More generally we can have mismatches leading to structures with near vanishing surface area and very large internal volumes, such as an inner expanding Universe that outgrew the outer one. The lifetimes of such structures can also be different, using the proper time of geodesic observers as a clock. We gave a number of concrete examples, but it is clear that an even larger class of solutions can be found.

Are these solutions realistic? They certainly could be part of the initial conditions of our Universe. But we can also speculate that they could be purely astrophysical objects: who knows what lurks out there?
We stress that the effects can be purely relativistic, so that interpreting, for example, the mass mismatch in a quasi-Newtonian framework (as the contribution of the kinetic or the binding energy) may not be appropriate. If a collapsing $k=-1$ OS-like model could be seen as the result of a violent non-gravitational implosion, it could also be a purely relativistic object. 

\section{Acknowledgments}
We thank Bernard Carr and Ray Rivers for discussions and Farbod Rassouli for assistance with the pictures. 
This work was supported by a Bell-Burnell Fellowship (RI) and partly supported by the STFC Consolidated Grants ST/T000791/1 and ST/X00575/1 (JM).



\appendix
\section{Israel Junction Conditions and the LTB Metric}\label{israel}
As explained previously, the LTB metric reduces to either the Schwarzschild metric in Lemaitre coordinates or the FRW metric. The outcome is contingent on one's choice of areal function $A(r,t)$, mass function $F(r,t)$ and curvature function $K(r)$. We may consider a scenario in which the Schwarzschild and FRW reductions of the LTB metric are glued at a common hypersurface. The two metrics partitioned by the hypersurface are therefore
\begin{equation}
    ds^2_{\pm}=-dt^2+X_{\pm}^2(r,t)dr^2+A_{\pm}^2(r,t)d\Omega^2.
\end{equation}
We take the hypersurface to be a surface of constant radius which is defined by the constraint $\Phi=0$ where $\Phi=r-r_{0}$. In general, a vector normal to the hypersurface defined by a constraint $\Phi$ is determined by the equation
\begin{equation}
    n_{\alpha}=\frac{\partial_{\alpha}\Phi}{|g^{\mu \nu}\partial_{\mu}\Phi\partial_{\nu}\Phi|^{\frac{1}{2}}}.
\end{equation}
For the LTB metric we arrive at
\begin{align}
    \begin{split}
        n_{\alpha}dx^{\alpha}=X(r,t)dr\\
        n^{\alpha}\partial_{\alpha}=\frac{1}{X(r,t)}\partial_{r}
    \end{split}
\end{align}
Generally, the projection tensors $e^{\mu}_{a}$ can be used to project out the components of a tensor which are normal to the hypersurface. In this notation, $x^{\mu}$ are coordinates associated with the full geometry while $y^{a}$ are the hypersurface coordinates. For example, the induced metric $h_{ab}$ is related to the full metric by the relation $h_{ab}=g_{\mu \nu}e^{\mu}_{a}e^{\nu}_{b}$. When two LTB metrics are glued together at a constant radius hypersurface, the induced metric on the hypersurface as approached from either metric is given by
\begin{equation}
    d\Sigma_{\pm}=-dt^2+A_{\pm}^2(r_{0},t)d\Omega^2,
\end{equation}
where $h_{tt}=-1$, $h_{\theta\theta}=A^{2}(r_{0},t)$ and $h_{\phi \phi}=A^{2}(r_{0},t)\sin^{2}\theta$. When dealing with metrics which are separated by hypersurfaces, we can define the following tensorial quantities;
\begin{align}
    \begin{split}
        &\big[h_{ab}\big]\equiv h_{ab(+)}-h_{ab(-)},\\
        &\big[n_{\alpha}\big]\equiv n_{\alpha(+)}-n_{\alpha(-)},\\
        &\big[k_{ab}\big]\equiv k_{ab(+)}-k_{ab(-),}
        \end{split}
\end{align}
where $\big[h_{ab}\big]$ is the difference between the induced hypersurface metric components across the hypersurface, $\big[n_{\alpha}\big]$ corresponds to a jump in the normal vector across the hypersurface and $\big[k_{ab}\big]$ is the difference in extrinsic curvature across the hypersurface. \\

\subsection{1\textsuperscript{st} Israel Junction Condition}
The 1\textsuperscript{st} Israel junction condition requires the geometry on the hypersurface should be the same when approached from both metrics. This corresponds to enforcing $\big[h_{ab}\big]=0$ across the hypersurface. In the case of two LTB metrics, enforcing the first Israel junction condition requires the identification
\begin{equation}
    \big[A(r_{0},t)\big]\equiv A_{+}(r_{0},t)-A_{-}(r_{0},t)=0.
\end{equation}
As mentioned previously, the traditional Friedmann-Schwarzschild junction is recovered under a specific choice of LTB free functions. The interior metric corresponds to Friedmann geometry when we choose the areal function to be 
\begin{equation}
A_{-}(r_{0},t)=a(t)r_{0}
\end{equation}
where $a(t)$ is the usual scale factor for the internal Friedmann region. We can identify the exterior geometry with the Schwarzschild metric in Lemaitre coordinates by choosing the corresponding areal function to be 
\begin{equation}
    A_{+}(r_{0},t)=\bigg[\frac{3}{2}(r_{0}-t)\bigg]^{\frac{2}{3}}r_{s}^{\frac{1}{3}},
\end{equation}
where the full LTB-Schwarzschild metric is given by
\begin{equation}
    ds^{2}=-dt^{2}+\frac{r_{s}}{A(r,t)}dr^{2}+A^{2}(r,t)d\Omega^{2}.
\end{equation}
In contrast to the usual curvature coordinates in which the Schwarzschild metric is usually expressed, Lemaitre coordinates are synchronous. Consequently, the time coordinate in the metric corresponds to to the proper-time of time-like observers. This choice of areal function can be interpreted as a surface which is initially at a maximum radius $r_{0}$ at initial proper-time $t_{0}$. Then for proper-time $t>t_{0}$, the surface is in free-fall from the maximum value $r_{0}$. In these coordinates, the analogy with the Friedmann scale factor is obvious. Indeed, the 1\textsuperscript{st} Israel junction condition is simply the interpretation of this free-fall in terms of the behaviour of the internal Friedmann scale factor $a(t)$.\\

It is worth noting that using the Schwarzschild metric in synchronous Lemaitre coordinates offers significant advantages over the usual curvature coordinates. In the usual gluing process, the internal Friedmann and external Schwarzschild metrics are taken to be

\begin{align}
    \begin{split}
        &ds^{2}_{-}=-dt^{2}+a(t)^{2}\bigg[\frac{dr^{2}}{1-kr^{2}}+r^{2}d\Omega^{2}\bigg],\\
        &ds^{2}_{+}=-f(r)dt^{2}+f(r)^{-1}dr^{2}+r^{2}d\Omega,\\
    \end{split}
\end{align}
where
\begin{equation}
    f(r)=1-\frac{2M}{r}.
\end{equation}
However in this case we cannot simply use the constraint on the $r$ coordinate for both metrics as we did in the LTB case. To match the internal Friedmann, geometry with the external Schwarzschild geometry, the radius of the surface as approached from the exterior Schwarzschild metric must be promoted to a function $R(t)$ where $t$ is Friedmann synchronous time. One must also take into account that the Schwarzschild time coordinate differs from synchronous time. The Schwarzschild time coordinate on the surface of the star is therefore promoted to $T(t)$, a function of the interior synchronous time. We may then take an induced metric on a $r=r_{0}$ hypersurface from the Friedmann perspective. This must be matched with the $r=R(t)$ surface from the Schwarzschild perspective. The induced metrics on the hypersurface are therefore given by

\begin{align}
\begin{split}
    &d\Sigma^{2}_{-}=-dt^{2}+a(t)^{2}r_{0}^{2}d\Omega^{2},\\
    &d\Sigma^{2}_{+}=-\bigg[f(R)\dot{T}^{2}+f(R)\dot{R}^{2}\bigg]dt^{2}+R^{2}d\Omega^{2}\\
\end{split}
\end{align}
Enforcing the 1\textsuperscript{st} Israel junction condition leads to the identifications
\begin{align}
    \begin{split}
        &-1=-\bigg[f(R)\dot{T}^{2}+f(R)\dot{R}^{2}\bigg],\\
        & R(t)=a(t)r_{0}.\\
    \end{split}
\end{align}
The first identification may be interpreted as matching the time on the surface of the dynamic Schwarzschild star with comoving Friedmann time. The second identification is simply an expression for the radius of the dynamic Schwarzschild star in terms of Friedmann variables.

\subsection{2\textsuperscript{nd} Israel Junction condition}
To enforce the 2\textsuperscript{nd} junction condition, we start with the definition of the extrinsic curvature $k_{\mu \nu}$ projected onto the the hypersurface $\Sigma$ with coordinates $y^{a}$;
\begin{align}
\begin{split}
    k_{ab}=&e^{\mu}_{a}e^{\nu}_{b}k_{\mu \nu},\\
    =&e^{\mu}_{a}e^{\nu}_{b}(\nabla_{\nu}n_{\mu}-a_{\nu}n_{\mu}),\\
    =&\nabla_{b}n_{a},
\end{split}
\end{align}
where $e^{\mu}_{a}$ is a projection tensor which projects out components of the extrinsic curvature normal to the surface $\Sigma$ as $e^{\mu}_{a}n_{\mu}=0$ and $a_{\nu}$ is the acceleration vector defined in terms of the normal as $a_{\nu}=n_{\lambda}\nabla^{\lambda}n_{\nu}$. As the hypersurface coordinates are $y^{a}=(t,\theta,\phi)$ and the LTB metric is diagonal, we're only concerned with the $k_{tt}$, $k_{\theta \theta}$ and $k_{\phi \phi}$ components of the extrinsic curvature. When considering the $k_{tt}$ component, it is worth noting that the LTB metric is both diagonal and expressed in a co-moving coordinate system. Therefore we would expect the $k_{tt}$ component of the extrinsic curvature to vanish
\begin{align}
\begin{split}
    k_{tt}= \ &\partial_{t}n_{t}-n_{r}\Gamma^{r}_{tt},\\
    = \ & 0,
\end{split}
\end{align}
since $\partial_{t}g_{at}=0$ for $a\neq t$ and $\partial_{r}g_{tt}=0$. The non vanishing components of the extrinsic curvature are therefore the angular components. Given that $\partial_{a}n_{r}=0$ for $a\neq r$ and $\Gamma^{r}_{ab}=-\frac{1}{2}g^{rr}(\partial_{r}h_{ab})$,
\begin{align}
    \begin{split}
        k_{ab}=&-n_{r}\Gamma^{r}_{ab}\\
        =&\frac{1}{2}n^{r}\big(\partial_{r}h_{ab}\big).
    \end{split}
\end{align}
The relevant components of the induced metric are $h_{\theta \theta}=A(r_{0},t)^{2}$ and $h_{\phi \phi}=h_{\theta \theta}\sin^{2}\theta$. Using the above formula for extrinsic curvature, the $k_{\theta \theta}$ and $k_{\phi \phi}$ are explicitly given by
\begin{align}
    \begin{split}
        &k_{\theta \theta}=A(r_{0},t) \cdot \frac{A'(r_{0},t)}{X(r,t)},\\
        &k_{\phi\phi}=k_{\theta \theta}\sin^{2}\theta.
    \end{split}
\end{align}
Enforcing the 2\textsuperscript{nd} Israel junction condition we have
\begin{align}
    \begin{split}
        \big[k_{a b}\big]=&\bigg[A(r_{0},t)\frac{A'(r_{0},t)}{X(r,t)}\bigg]\\
        =&\bigg[\frac{A'(r_{0},t)}{X(r,t)}\bigg]\\
        =&\big[K(r)\big].
    \end{split}
\end{align}
The second line is derived from the first by using the fact that we've enforced the first Israel junction condition that $\big[A\big]=0$. The third line is then a consequence of the momentum constraint for the general LTB metric leading to the identification $X=C(r)A'$ where $C(r)=1/\sqrt{1-K(r)}$ with $K(r)<1$.

\end{document}